\documentclass{article}
\input epsf

\newcommand{\eqn}[1]{(\ref{#1})}
\def\be{\begin{equation}} 
\def\ee{\end{equation}}
\def\bea{\begin{eqnarray}}
\def\eea{\end{eqnarray}}
\def\nn{\nonumber}
\def\secnum#1{\section{#1}\setcounter{equation}{0}}
	
\font\smallrm=cmr6

\def\br{\hfil\break}
		
          \def\b{\beta}   \def\g{\gamma}  
\def\d{\delta}            
            \def\k{\kappa}  \def\l{\lambda}          
\def\m{\mu}             \def\f{\phi}    \def\F{\Phi}            
\def\n{\nu}             \def\j{\psi}    
\def\r{\varrho}         \def\s{\sigma}  \def\SS{\Sigma}
            \def\th{\theta}      
                     
\def\w{\omega}  \def\W{\Omega}                          

  \def\OO{{\cal O}}

\def\cl{\centerline}    
      \def\pa{\partial}       \def\dd{{\rm d}}        
\def\tl{\tilde}                 \def\bra{\langle}       \def\ket{\rangle}

\def\ffract#1#2{{\textstyle{#1\over#2}}}
\def\fract#1#2{{\raise .35 em\hbox{$\scriptstyle#1$}\kern-.25em/
	\kern-.2em\lower .22 em \hbox{$\scriptstyle#2$}}}

\def\half{\fract12} 

\def\part#1#2{{\partial#1\over\partial#2}}   
\def\ex#1{e^{\textstyle#1}}

\def\bbf#1{\setbox0=\hbox{$#1$} \kern-.025em\copy0\kern-\wd0
        \kern.05em\copy0\kern-\wd0 \kern-.025em\raise.0433em\box0}              

\def\low#1{{\vphantom{]}}_{#1}} \def\up#1{{\vphantom{]}}^{\textstyle {#1}}}

\def\Gbar{\raise.13em\hbox{--}\kern-.35em G}
 \def\ins {\low{\rm in}}\def\out{\low{\rm out}}
\def\ra{\rightarrow} \def\haw{\low{\rm h}}

\title{Distinguishing causal time from Minkowski time and\break
a model for the black hole quantum eigenstates}
 
\author{G. 't Hooft 
  \footnote{Institute for Theoretical Physics,
 Utrecht University, \br
 \hbox{\ \ \ } Princetonplein 5,
 3584 CC Utrecht, 
 the Netherlands. \br
 \hbox{\ \ \ } email: g.thooft@fys.ruu.nl .}}
  
\begin{document}

\maketitle
\cl{THU-97/36 \hfil  gr-qc/9711053}
\abstract{A discussion is presented of the principle of black hole
complementarity.  It is argued that this principle could be viewed as
a breakdown of general relativity, or alternatively, as the
introduction of a time variable with multiple `sheets' or
`branches'.\par A consequence of the theory is that the stress-energy
tensor as viewed by an outside observer is not simply the
Lorentz-transform of the tensor viewed by an ingoing observer.  This
can serve as a justification of a new model for the black hole
atmosphere, recently re-introduced.  It is discussed how such a model
may lead to a dynamical description of the black hole quantum
states.}

\secnum{Introduction} 

The gravitational force, as dictated by General Relativity, has a
built-in instability.  If a sufficiently large quantity of matter is
concentrated in a sufficiently small volume, collapse is inevitable,
and a black hole results.  It is characterized by a space-time in
which an event horizon occurs, as is sketched in Fig.~1a.  The
horizon is depicted as a transparant surface; it is a lightlike surface defined in
such a way that any material object inside it, is unable to escape
from this region as long as its velocity force vector stays
within its local lightcone.

From the point of view of classical physics, one can predict precisely
what will happen, and there seems to be no serious clash with astronomical
observations.  However, when one wishes to study the consequences of
quantum field theory in these circumstances, new difficulties arise that
are intensively being studied.  It was discovered by Hawking \cite{H1} that
quantum field theory will force the black hole to emit particles, with the
intensity of a thermal heat source at temperature 
\be T_H={\hbar c^3\over 8\pi GM}\,,\label{TH}\ee
where $M$ is the mass of the black hole.  Details
of the derivation of this effect can be found in refs. \cite{H1,H2,H3,GtH2}

The fact that the radiation emitted, as described by \eqn{TH}, is {\it
thermal}, opens up the possibility to approach this phenomenon from a
thermodynamical point of view. \cite{B1}. Taking $M$ to be the energy 
(using units in which $\hbar=c=1$), and $T=T\low H$ the
temperature, one readily derives the {\it entropy}~$S$: 
\be T\dd S\,=\,\dd M \,;\qquad\dd S\,=\,8\pi G M \dd
M \,;\qquad S\,=\,4\pi G M^2 +C\,,\label{entropy}\ee
where $C$ is an unknown integration constant, to be referred to as the
``entropy normalization constant". 
It is important to note that the expression obtained for the entropy $S$,
apart from the integration constant, is always equal to $\ffract14  A/G$, where
$A$ is the {\it area} of the horizon, a finding that will be very much at the
center of our discussions.  A physical interpretation of this entropy is that
{\it information\/} appears to be distributed over the 2-dimensional area
of the event horizon in such a way that one bit of information occupies
an area exactly amounting to
\be 4G\ln 2\ =\ 0.724\times 10^{-65}\,\rm cm^2\,.\label{bit}\ee

\epsffile{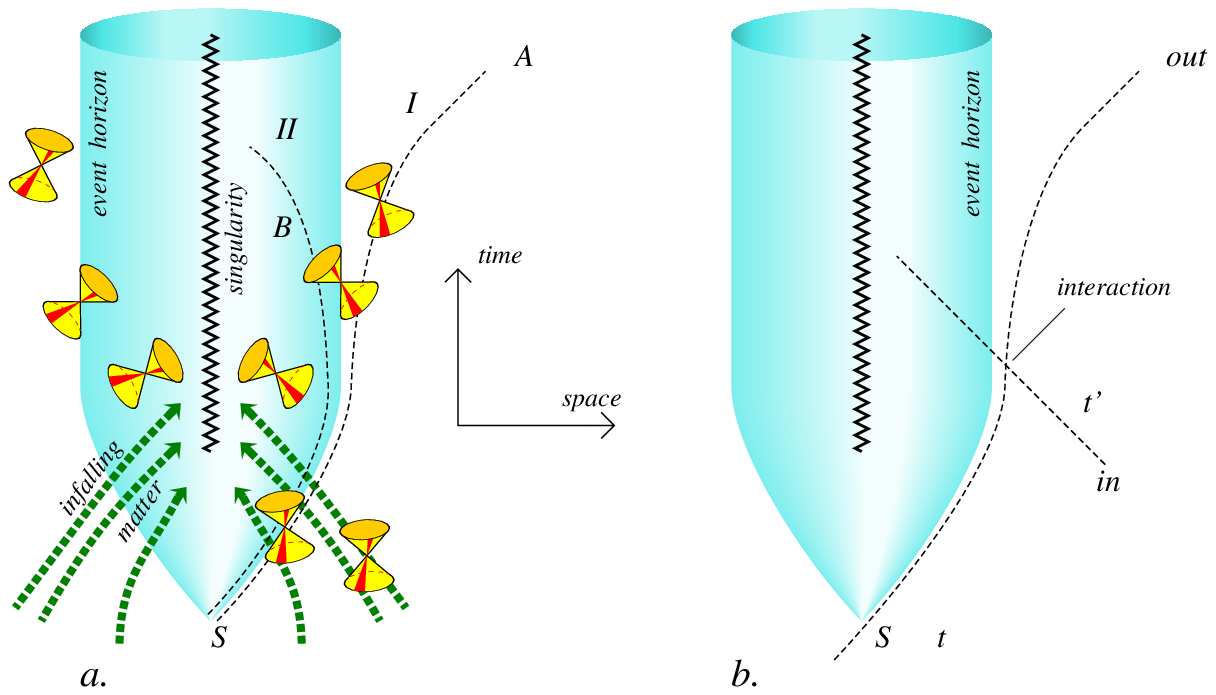}
\cl{Fig.~1. Space-time sketch of gravitational collapse.}
\bigskip 

It is tempting to assume that a black hole can exist in $e^S$ distinct 
quantum states, and, apart from the fact that this number can take huge
values when the hole is large, these quantum states each can be compared with
ordinary elementary particles. Furthermore, one would expect that the entire
process of formation and subsequent desintegration of a black hole can be
described by a scattering matrix, as in quantum scattering theory.

In the next section, however, it will be explained that this simple assumption
appears to lead to conflicts; a genuine paradox is encountered. It is
called the `quantum information problem', and it is believed to have
a deep and fundamental significance for the theory of quantum gravity.

The most promising avenue towards a resolution of the paradox is what is
called the `$S$-matrix Anstz', as explained in Sect.~3, but this does
lead to the new and peculiar feauture of `complementarity' (Sect.~4).
In Sect.~5 we introduce a multivalued time parameter.

The second half of this paper is a description of a model based on these
ideas. The outside observer here treats the Hawking particles he/she
observes as genuine sources for the Einstein equations, in spite of the fact
that ingoing observers would disagree. Although an unphysical singularity
emerges at the origin ($r=0$), the model allows for a detailed
description of the black hole quantum states. It does appear to require an
exotic equation of state for matter in the Planck regime.

\secnum{The quantum information problem}

Imagine the black hole formation process, assuming that the collapsing object
started in a pure state. The wave packet of a particle moving along the 
trajectory labled $A$ in Fig.~1a, enters the black hole region near the  
pint $S$ (where the horizon opens up). There, it
corresponds to a pure state $|\j\ket\ins =|\W\ket$. During its evolution,
this wave packet will split into two parts, one leaving the hole near the point
$A$ in region $I$, and one entering the hole in region $II$.
This state thus evolves into a superposition of product states:
\be|\j\ket\ins\ =\ \prod_{\hbox{\smallrm modes}\atop\w}\sqrt{1-\ex{-8\pi M\w}}\,
\sum_n\ex{-4\pi M\w n}|n\ket\low I|n\ket\low{II}\ \ 
=\ \sum_i|\j_i\ket\low I|\j_i\ket\low{II}\,,\label{modes}\ee 
the last line being short-hand notation. $n$ stands for the 
number of particles at a given frequency~$\w$.

In terms of observable operators $\OO$ that act only upon the visible 
states $|\j_i\ket_I$, one finds that this state is a {\it quantum mechanically
mixed\/} state:
\be\bra\OO\ket\ =\ \sum_{ij}\bra\j_i|\low I\OO\r_{ij}|\j_j\ket\low I\ ; \qquad
\r_{ij} \ =\  {}\low{II}\bra\j_i|\j_j\ket\low{II}\ =\ {\rm Tr}\r\OO\,.\ee 
Here, the matrix $\r$ is a quantum mechanical density matrix, and, since
in Eq.~\eqn{modes}, the quantity $\w n$ is the energy of the state $|n\ket$, one
easily derives from \eqn{modes} that this density matrix is a thermal one,
corresponding to the Hawking temperature \eqn{TH}. 
 
Now, imagine an amount of matter in a quantum mechanically {\it pure\/}
state, undergoing gravitational collapse.  The state $|\j\ket\ins $ can be
taken to be pure as well.  Then, if the states $|\j\ket_I$ are taken to be
the only ones accessible to the outside observer (after some time has
elapsed), one would have to conclude that the final state is a mixed 
state \cite{H4}.
Did an evolution take place where a pure state evolved into a mixed state?
This would be against the rules of conventional quantum mechanics.
Alternatively, should we keep the states $|\j\ket_{II}$ in our Hilbert
space forever?  In that case, one deduces that the dimensionality of
Hilbert space would quickly grow much larger than suggested by the rather
low value \eqn{entropy} of the total entropy --
the level density would become a continuum, instead of describing a
denumerable, discrete set.  Because of the infinite number of possible
states, one would argue that such black holes would obey Boltzmann
statistics instead of Bose-Einstein or Fermi-Dirac statistics, which is the
law for ordinary matter.  There would be no such thing as a scattering matrix.

From a physical point of view, one expects that the density matrix is
merely approximately thermal, but that true thermality only occurs at the
infinite size limit.  The {\it $S$-matrix Ansatz\/} \cite{GtH1, GtH2} 
asserts that Hawking
radiation is actually governed by an $S$ matrix. The entropy \eqn{entropy}
would indeed refer to the total number of possible states, and the 
{\it quantum information\/} of all particles that enter
into the hole during its entire lifetime, does not disappear, but reemerges
with the Hawking radiation.  More precisely, if we compare two initial
states that are orthogonal to each other (which already is the case if only
one of its ingoing particles has its spin reversed), then the two
corresponding pure out-states will be orthogonal as well, according to the
laws of unitary evolution.
 
Consider Fig.~1b.  Imagine a particle with spin up, entering the hole
at late time $t'$.  If we flip its spin, the outgoing particles, seen at
all points $A$, should change into a new quantum state, orthogonal to the
previous one.  The paradoxical aspect of this, however, is that these
outgoing particles {\it all\/} originate at the point $S$ where the horizon
first opened up.  The change should therefore already be visible at that
point, but $S$ is located at time $t$ much earlier than $t'$.  Thus, the
$S$-matrix Ansatz appears to violate causality.

In Fig. 1b, one can also see how causality can be safeguarded. We need to
consider the {\it interactions} that take place where the ingoing particles
meet the outgoing ones. In a linear theory, {\it i.e.} in the absence of
interactions, there would be no causality. But the interactions cannot
be ignored. Most of all, the {\it gravitational\/} interactions between
in- and outgoing material is seen to diverge with the time difference
$t'-t$.

Another way to see the difficulty for linearized theories, is to construct the equations
for quantum fields in the Schwarzschild geometry near the horizon. After
collapse, this space-time is described by the metric
\be\dd s^2=-\left(1-{2M\over r}\right)\dd t^2+{\dd r^2\over
1-2M/r}+r^2\dd \W^2\,,\label{SS}\ee where Newton's constant $G$ was normalized
to one, and 
\be\dd\W^2\equiv\dd\th^2+\sin^2\th\dd\f^2\,.\ee
Working out the field equations, one finds it to be convenient to
substitute
\be r-2M=\ex\s\ ,\qquad -\infty<\s<\infty\,.\ee
A free scalar field at angular momentum $\ell$ then obeys
\be\left[r\Big(\part{ }t\Big)^2-{1\over r^2}\part{ }\s r\part{ }\s
+\ex\s\left({\ell(\ell+1)\over r^2}+m^2\right)\right]\Phi=0\,.\ee
At $\s\ra-\infty$, this produces plane waves moving in and out, but
ingoing waves do not scatter back in any finite amount of time; hence
a strictly continuous spectrum of states results, which is at odds with
the finiteness of the entropy. Of course, gravitational interactions
diverge as $\s\ra\infty$.

If a {\it brick wall\/} would be introduced \cite{GtH3}, with which we mean a
reflecting boundary condition at some small but finite value of $\s$, a
discrete spectrum could be enforced.  We do notice that such a brick
wall would correspond to a reflecting envelope at a small but finite
distance away from the horizon in Fig.~1b.  Clearly, it would be enough te
reestablish an $S$-matrix but it is clear that such a boundary condition
would violate general coordinate invariance. Or could it be that the divergent
gravitational interactions mimick a brick wall?

Why is the information problem considered to be so important?  First of
all, it appears that what is at stake here is the extent of {\it
predictability\/} of the fundamental dynamical processes at the Planck
length.  Some theories of the Universe \cite{linde} suggest that in many --
extremely distant -- regions of the Universe, different laws of physics
apply; for instance, the unifying gauge symetry may condense into gauge
subgroups different from $SU(3)\times SU(2)\times U(1)$, which we are used
to in the Standard Model.  Is there indeed nothing more than the anthropic
principle that determines the symmetry breaking and the associated
constants?  If so, this will mean that we will never be able to compute
these effects from first principles more refined than this anthropic
principle.

Of even more importance, however, appears to be the fact that {\it
assuming\/} the preservation of quantum information leads to interesting
new insights in the forces of nature.  Conservation of quantum information
is likely to demand a new kind of conspiracy, and the resolution of the
paradox may well lead to important new physics.  Indeed, we see examples of
important new developments in physics in the past, that sprouted from
attempts to resolve paradoxes.

\secnum {The $S$-matrix Ansatz}

At first sight, it seems to be a necessary consequence of the imput theories,
being quantum mechanics and general relativity, that the outgoing state is
related to the ingoing state through a well-defined linear mapping, called
the $S$-matrix. Yet, as was argued in the previous section, this appears
to be not true if the laws are taken at face value and the calculation is
performed. In this section, we {\it postulate\/} \cite{GtH1} that 
indeed there is an
$S$-matrix. When this postulate is taken together with the known laws of
physics, new features can be derived.

We begin by first concentrating on a small region of the black hole
horizon, ignoring the local curvature of space-time. One then gets
``Rindler space'', pictured in Fig.~2a. The frame used to draw the picture
is the one in terms of which the metric stays regular naer the horizon.
In terms of the original coordinates $r$ and $t$, the Schwarzschild 
metric~\eqn{SS} is singular at $r=2M$. A few lines $t=$~constant and
$r=$~constant, are indicated. A translation in time, $t\ra t\,+$~const.,
corresponds to a Lorentz transformation in terms of regular 
coordinates.

\epsffile{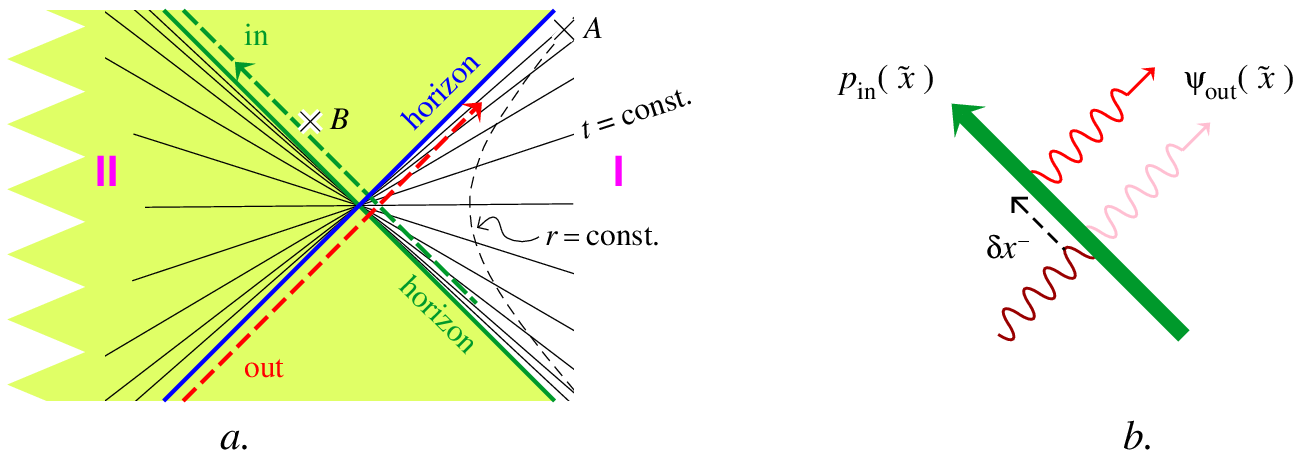}
{\narrower{Fig.~2. a) Rindler space. The unshaded 
region is the visible part of the universe. b) The gravitational shift 
of an outgoing particle due to
the gravity of an ingoing one.}\par}\bigskip

According to the $S$-matrix Ansatz, it is to be expected that the ingoing
particles (dotted line from lower right to upper left) {\it affect\/}
the outgoing particle state (dotted line from lower left to upper right).
This interaction must occur at the intersection point (which is actually a
two-dimensional plane since we still have the angular variables $\th$ and
$\f$. The interactions due to the forces of the Standard Model are weak,
and it is
hard (but not impossible) to take these interactions into account. But the
gravitational interactions are much more profound. Since time translations
for the distant observer correspond to Lorentz transformations for a
local observer, the late outgoing particles cross the trajectories of
early ingoing ones at very large relative velocities. It is here where the
gravitational interactions become important -- growing as they do with energy.

The gravitational field due to a very fast particle, with small rest mass, can
be computed by standard methods of general relativity. One finds that the 
field of an ingoing particle affects the outgoing ones by giving them a
{\it shift} \cite{DGtH}. This is pictured in Fig.~2b.

The shift $\d x^-$ in the coordinate $x^-$ for the outgoing particle depends
on the transverse separation $\d\tl x$ between the two particles:
\be\d x^-(\tl x) = \int\dd^2 \tl x'f(\tl x-\tl x')\,p\ins (\tl x')\,,\ee
where the momentum $p\ins $ is taken as a distribution over the transverse
(angular) coordinates $\tl x=(\th,\,\f)$. The function $f$ has a logarithmic
singularity where $\tl x-\tl x'\ra 0$. Its form is dictated by Einstein's 
field equations.

Our general philosophy, exhibited in much more detail in Ref. \cite{GtH2}, is as
follows. Consider a black hole, formed by the collapse of a large amount of
matter. It will produce outgoing particles in a set of very wide wave packets
$\j\out(\tl x')$. Now, consider an infinitesimal {\it change\/} in the
quantum states of the ingoing particles. The distribution $\d p\ins (\tl x)$
will undergo a tiny change, $\d p\ins (\tl x)$. This causes a corresponding
shift in the outgoing states, described by a quantum mechanical shift
operator. Thus, indeed, we find that the outgoing state depends on the ingoing 
one. The most delicate part of the argument is then the requirement that
this dependence be described by a unitary scattering matrix operator.
A new version of Fock space must be introduced, and what is found shows
a startling resemblance to string theory amplitudes -- though it is not
exactly string theory that we find.\footnote{The {\it string coupling constant\/}
is found to be purely imaginary.}

\secnum {Black hole complementarity}
 
Although, in principle, the theory outlined above should reveal the nature
of the quantum states on the black hole horizon, the paradox mentioned
earlier is not entirely resolved.  The pivotal question is, how does an
ingoing observer experience and identify these quantum states, and how can
we reconcile the finiteness of their density with Lorentz invariance?
Should Lorentz transformations not lead to strictly infinite numbers of
states?  And, secondly, how is the apparent clash with causality diverted?

The latter question can be phrased more precisely.  Consider Rindler space
as it is sketched in Fig.~2a.  Consider operator-valued
fields in the Heisenberg picture.\footnote{The points $A$ and $B$ 
in Fig.~2a  more or less correspond
to $A$ and $B$ in Fig.~1a.} Thus, consider the point $B$ in Fig.~2a,
and fields $\F(x)$ with the space-time point $x$ in the immediate
neighbourhood of $B$.  Its operators can only act on the states described by
the early ingoing particles, but all visible outgoing particle wave packets
are well outside the lightcone defined by $B$, and so one might conclude
that $\F(x)$ should commute with {\it all\/} operators $\F(y)$ with $y$ in
the direct neighborhood of the point $A$, where an observer detects Hawking
particles.  If, however, the Hawking particles, all of which can be detected
near the point $A$, would form a complete representation of Hilbert space,
no operator acting on the ingoing particles could commute with all Hawking
operators at $A$.

According to the $S$-matrix Ansatz, the operators at $A$ will therefore {\it
not\/} all commute with the operators at $B$.  The reason why the light cone
argument breaks down is the excessive relative energies of the particles in
consideration; due to these energies, the space-time metric is distorted,
and the light cone will not always stay in position.

Again consider an operator field acting near the point $B$.  What is really
meant by this field operator is that it acts in a certain way on the ingoing
states, and this action can be derived by extrapolating the fields of all
ingoing particles well into the forbidden (shaded) region of Fig.~2a.  In
making this extrapolation, it is tacitly assumed that the penetration
into this region is unproblematic, since space and time are free from
excessive curvature near the horizon.

Now, however, consider an observer performing a measurement near the point
$A$.  This observer detects Hawking particles.  If he waits long enough, his
time coordinate $t$ is so large that the particles he detects all carry
enormous amounts of energy with respect to the Lorentz frame of $B$.  The
action of $A$ on any state in Hilbert space, in general will produce a state
in which $B$ does not see a near vacuum at the origin, but one or more
extremely energetic particles there.  These particles carry devastating
gravitational fields.  The ingoing particles are, in fact, deflected by
these fields.  The {\it tacit assumption\/} of the previous paragraph, that the
operator in $B$ could be defined as if the ingoing particles could reach
this point undisturbed, has become untenable.  This is why the commutators
between fields at $A$ and fields at $B$ cease to vanish.

In fact, if observers are allowed to act on our states at sufficiently late
points $A$, any attempt to define the action of fields at $B$ will become
meaningless.  We have to {\it choose\/}:  either we allow the operators
defined at $A$, or the operators defined at $B$, but not both sets.  Black
hole complementarity \cite{STU} is the principle that a {\it complete\/} 
description of
Hilbert space can be obtained either in the frame of the ingoing observer,
allowing him to do measurements at points $B$, or by allowing all
measurements on the emitted Hawking particles.  This is an inevitable
consequence of the $S$-matrix Ansatz.  It also delivers a picture that
appears to be more compatible with time reversal invariance than the
classical picture of black holes losing information.

\secnum {The forked time paramter} 
 
The linear mapping in the previous section, relating $A$ and $B$,
may be quite complicated. This is in contrast to the more usual
general coordinate transformations, for which the transformation rules
are direct and transparent. Indeed, the mapping will probably be as
difficult to elaborate as the time evolution itself. This may be tantamount
to saying that the general coordinate transformation itself is not
able to provide us with the information we want about the evolution of these
states, or in other words, the general coordinate transformation fails.
The transformation works if we want to handle the states $|\j\ket_I$
and $|\j\ket_{II}$, but not for the states $|\j\ket\ins$ or
$|\j\ket\haw$. Although the detailed description of the states seen by an
observer going into the hole, does contain information concerning the
phenomena seen by the late outside observer, this information is thoroughly
encrypted.
 
\epsffile{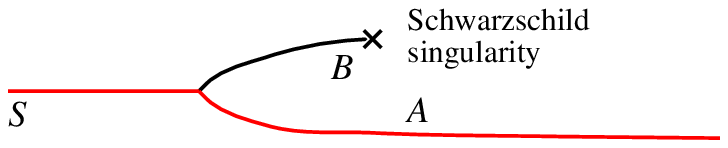}
\cl{Fig. 3.  The multi-valued time parameter.}\bigskip
 
In Fig.~3, this is further illustrated. The collapse is observed at the
point $S$. The ingoing observer follows the path $B$ towards the
Schwarzschild singularity. The outside observer continues along the path
$A$. After the collapse, Hilbert space is {\it not\/} described by the
products of the states along path $A$ and the ones along path $B$, 
but it is spanned either by the states at $A$ alone, or by the states
at $B$ alone. The Hamiltonian $H$ is defined along $A$ and along $B$,
and it dictates the evolution everywhere along the curve. Thus, we see
that time has become a manifold more complicated than just a single line.

An interesting implication of this view is that the {\it stress-energy-momentum
tensor\/} $T_{\m\n}(x)$ as seen by $B$, will be defined differently from
the one seen by $A$. Energy, after all, is related to the time translation
operator, but translation in the $B$ direction is now no longer
equivalent to translation in the $A$ direction. Therefore, it has
become illegal to relate $T_{\m\n}$ as seen by $B$ to that of $A$ through
the usual coordinate transformations. Now, usually, it is assumed
that $B$ will detect a vacuum state ($T_{\m\n}\approx0$). We are not
allowed to infer from this that $A$ will see a cancellation between
the stress-energy-momentum of the Hawking particles against a large (and negative)
contribution from the Casimir effects in the Schwarzschild metric, as
was reported in earlier literature \cite{DFU,UW}. 

This gave us the inspiration to perform the calculation sketched in the next 
sections. The stress-energy-momentum tensor as experienced by an
observer of Hawking radiation is {\it not\/} considered to be the Lorentz
transform of an empty space stress-energy-momentum tensor. We allow it to
be the full tensor of genuine matter. This simply means that the subtraction
of the vacuum contribution is not done as in Ref. \cite{DFU}. The conformal
anomaly obtained by \cite{DFU} is here regarded as an anomaly in the
general coordinate transformation itself. In some sense, one may state
that general coordinate invariance is violated. The violation arises from
the fact that Hilbert space, as viewed by an ingoing observer, does not
contain the same states as the one handled by the outside observer.
The cut-offs are performed differently.

\secnum{A model}
 
For studying the complete set of quantum states of a black hole, it is
not sufficient to limit oneself to just the single case where ingoing
observers see only empty space, the so-called Hartle-Hawking state.  
The modes to be studied in
this paper are stationary in time.  In some sense, they are unphysical.
This is because a black hole in equilibrium with an external heat bath
cannot be stable; it has a negative heat capacity, and it is not
difficult to deduce that thermal oscillations therefore diverge.  Thus,
the states that will be discussed in this paper, being stationary in
time, are good candidates for the quantum microstates, but do not
represent the Hartle-Hawking state.  This way we claim to be able to
justify the calculations carried out in this paper:  we simply take the
stress-energy-momentum tensor generated by Hawking radiation as if the Hawking
particles represented an `atmosphere'.  This atmosphere is taken as the
source for the gravitational field equations.  We ignore its
instability, which is relevant only for much larger time scales.
Thermal, non-interacting, massless particles obey the equation of state,
\be p=\fract13\r\,, \label{onethird}\ee  
where $p$ is the pressure and $\r$ the energy density.  If we have $N$
massless (bosonic) particle types, at temperature $T=1/\b$, we have
\be \r=3p={\pi^2\over 30}{N\over\b^4}\,.\label{init}\ee  

In this paper, we regard Eq.~\eqn{onethird} or~\eqn{init} as 
describing matter near
the black hole, and at a later stage we will substitute the true Hawking
temperature for $\b^{-1}$.   Taking Hawking radiation as
a description of the boundary condition at some large distance from the
horizon, we continue the solution of Einstein's equations combined with
the equation of state as far inwards as we can.  What is found is that
Hawking radiation produces radical departures from the pure
Schwarzschild metric near the horizon.  In fact, the horizon will be
removed entirely, but eventually a singularity is reached at the origin
($r=0$).  This should have been expected; our system is unphysical in
the sense that, when $M$ is sufficiently large, the Shandrasekhar limit
is violated, so that a solution that is regular everywhere, including
the origin, cannot exist.

We must mention that the differential equations
for this system, the so-called Tolman-Oppen\-hei\-mer-Volkoff
equations \cite{TOV}, have been studied before by Zurek and Page \cite{ZP}, and
although they use slightly different variables, many of their conclusions were
identical to the ones rederived by the author \cite{GtH4}.  
 
The negative-mass singularity is the only way in which this approach
departs (radically) from earlier Einstein-matter calculations \cite{Bi}.  But
for the remainder of our considerations this singularity is harmless.
It being repulsive, all particles will keep a safe distance from this
singularity.  This then, enables us to compute quantum states in the
metric obtained.  Thus,  we compute the
entropy due to the scalar fields. We
find that this entropy is finite, so that the Hawking `blanket' itself
apparently acts as a soft alternative to the `brick wall' introduced in
Ref. \cite{GtH3}.  Furthermore, we find  the total entropy of all particles to be
independent of $N$, but to slightly overshoot Hawking's value.
Presumably, the equation of state~\eqn{onethird}
 was too much of a simplification.

It so happens that the metric can also be calculated if Eq.~\eqn{onethird} is
replaced by \be p=\k\r\,,  \ee  where $\k$ is a coefficient ranging
anywhere between 0 and 1.  Moreover, the total entropy can also be
calculated in this case.  One may decide to adjust the value of $\k$
such that the entropy calculation matches precisely, but this could be
premature, because here one cannot ignore the quantum corrections to
Einstein's equations, since we are operating in the Plankian regime.  It
may nevertheless be of interest to note that a complete match is
achieved if $\k$ is set equal to 1, which is the case that will be
elaborated further in the Appendix.  It is a rather singular and
unphysical case.  We do conclude that, with interactions taken into
account, it may well be possible to obtain a self-consistent approach
towards microcanonical quantization of black holes, using ordinary
Hartree-Fock methods in standard gravity theories.  We do stress that
the price paid was a (mild) singularity at the origin.  A more thorough
analysis of the exact role played by this singularity in a more
comprehensive theory of quantum gravity is still to be performed.
 
\secnum{The equations for $\k=\ffract13$} 
 
We only consider spherically symmetric, stationary metrics in $3+1$
dimensions, of the form
\be \dd s^2=-A(r)\dd t^2+B(r)\dd r^2+r^2\dd\W^2\,, \label{metric} \ee 
having as a material source a perfect fluid with pressure $p(r)$ and
energy density $\r(r)$. The Einstein equations read\footnote{Here, 
units are chosen such that $G=1$.}
\be  1-\pa_r\left(r\over B\right) = 8\pi r^2\r\,,\qquad
{\pa_r (AB)\over AB^2r}=8\pi(\r+p)\,, \label{Ein}\ee 
where $\pa_r$ stands for $\pa/\pa r$.  The relativistic Euler
equations for a viscosity-free fluid can be seen here to amount to \cite{LF}
\be \pa_rp=-(\r+p)\pa_r\log\sqrt{A(r)}\,,\label{euler}  \ee 
and the relation between $p$ and $\r$ is governed by an equation
of state. We will consider the case
\be p=\k\r\,, \label{kappa} \ee 
where $\k$ is a fixed coefficient.  Massless, non-interacting particles
in thermal equilibrium have $\k=\ffract13$.  The calculations
described here can be performed for any choice of $\k$ between $0$ and
$1$, but for brevity we concentrate on the special choice $\k=\ffract13$.
  For the general case we refer to Ref. \cite{ZP, GtH4}.
In particular, if $\k=1$,  complications arise \cite{GtH4}.  
Our liquid will be viscosity-free and free of vortices, 
so that Eq.~\eqn{euler}  can be integrated to yield 
\be 8\pi\r A^2=C\,,\label{asympt} \ee  
where $C$ is a constant, later to be called $3\l^2/(2M)^2$.  After
inserting this equation into Eqs~\eqn{Ein} --\eqn{kappa} , the latter can be cast in
a Lagrange form, which will not be further discussed here.  What is needed is
to observe the scaling behavior as a function of $r$.  Scale-independent
variables are $X$ and $Y$, defined by
\be X=C^{-\half} A\,r^{-1}\,,\qquad
\hbox{and}\qquad Y=B\,.\label{XY}\ee 
This turns the equations into:
\be  {r\pa_rX\over X}={Y\over 3X^2}+Y-2\ ,\qquad
{r\pa_rY\over Y}=1-Y+{Y\over X^2}\,.\label{eqXY}\ee  
Eliminating $r$, yields a first order,
non-linear differential equation relating $X$ and $Y$.

The result of a numerical analysis of this equation is presented in Fig.~4.
For large $r$, all solutions spiral towards the
point $\W$.  For small $r$, only one curve allows $B$ to approach the value
one, so that, according to Eq.~\eqn{Ein} , the density $\r$ stays finite, and
the metric~\eqn{metric} stays locally flat.  This is the only regular solution.
Further away from the origin, this solution requires $\r$ always to be
large.

 \cl{\epsffile{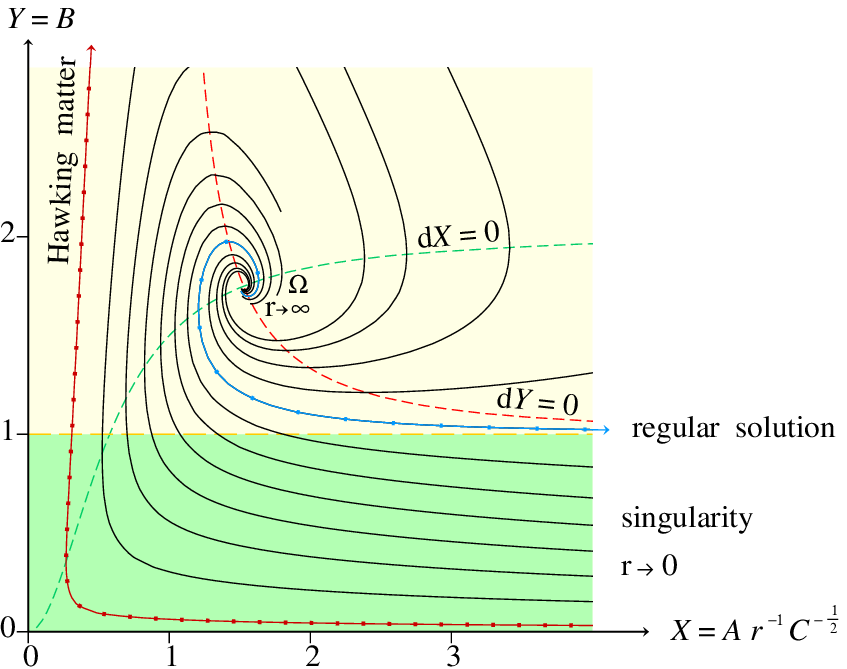}}
\cl{Fig. 4. Solutions to Eqs. (2.9) -- (2.10).}\cl{Solutions entering the 
 shaded region must have a singularity at $r=0$.} \bigskip 

We are interested in a very different class of solutions, the ones which,
far from the origin, approach a black hole surrounded by a very tenuous
cloud of matter, Hawking radiation.  This is the case where $Y\approx1$ 
and $X\gg1$. Observe that, for very large
$r$, all solutions will eventually spiral into the point $\W$:
\be \W=\left(X=\sqrt{7C/3}\,,\ Y=\ffract74 \right)\,. \ee  
Physically, this means that, since the universe is actually filled with
radiation, the curvature at large distances becomes substantial. For the
study of Hawking radiation, this large distance effect is immaterial and
will henceforth be ignored.

Thus, our boundary condition far from the origin will be chosen
to be
\be A(r) = 1-{2M\over r} \ ,\qquad
 B(r)=\left(1-{2M\over r}\right)^{-1}\ ,\qquad
 8\pi\r(r) ={3\l^2\over(2MA)^2}\ ,\label{er1}\ee 
in the region \be \l\ll {r\over 2M}-1\ll {1\over\l}\,. \label{r1}\ee 

Here,  the constant $C$ was replaced by $3\l^2/(2M^2)$,
since $\r$ has dimension $(\hbox{mass})\times(\hbox{length})^{-3}$
$=\,(\hbox{mass})^{-2}$, and the factor 3 is for later convenience.
  $\l$ will be dimensionless:
\be  \l^2={4\pi^3\over45}{N\over\b^4}(2M)^2\,,  \ee 
where $\b$ is the inverse temperature. 
In Planck units, Hawking radiation has
\be \b=8\pi GM=8\pi M\qquad \ra\qquad  2M\l={1\over24}\sqrt{N\over5\pi}\,,
\label{lambda} \ee 
and hence for large black holes, $\l$ is very small. 
The approximation~\eqn{r1} holds over a huge domain.

When $r$ approaches the horizon, the effects of $\r$ are nevertheless
felt, and the solution becomes more complicated. Actally, $A$ never
goes to zero, so there is no horizon at all. At small $r$,
$A$ diverges as a $\hbox{constant}/r$, which means that there is a
negative-mass singularity.

Following the line of the solution of interest in Figure~4, one observes
 that, for small enough $\l$, the solution can be found
analytically. It will be elaborated in the next section. 
 
\secnum{The solution for small $\l$} 
 
Eqs.~\eqn{eqXY} become slightly easier if we substitute $X$ and $Y$
by $P$ and $Q$:
\be P=3X^2/Y\ ;\qquad Q=1/Y\ ;\qquad
\hbox{hence}\qquad X=3^{-\half}\, P^\half\, Q^{-\half}\,. \ee  
Defining $L=\log r$, the equations become
\be  \part PL={3P\over Q}-5P-1\ ;\qquad
\part QL=-{3Q\over P}-Q+1\,.\label{PQ}\ee  
In all limiting cases of interest to us, the r.h.s. of these equations will 
simplify sufficiently to make them exactly soluble.

In order to integrate these equations down towards $r\ra 0$, 
we have to glue together different regions, where different approximations are
used. All in all, we cover the line $0<r\ll{2M\over\l}$ with six
overlapping regions, as depicted in Figure~5 (in the $\k=1$ case, these
will be 8 regions).

\cl{\epsffile{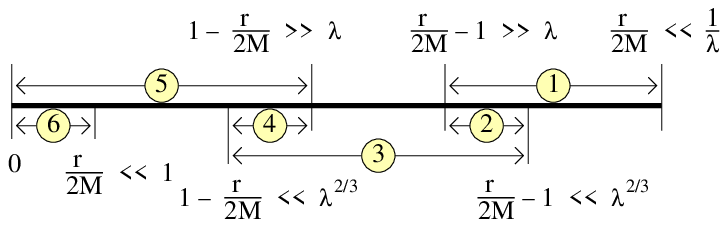}}
 
\cl{Fig. 5. Covering the line $0<r\ll 1/\l$ with regions (1) -- (6).} 
\bigskip

The region of Eq.~\eqn{r1} is region~(1). The effects of $\r$ on the metric are
negligible, and so $A$ and $B$ still obey Eqs.~\eqn{er1}. These
expressions agree with Eqs.~\eqn{PQ} if we take
$P\gg Q$.

In region~(3), a different approximation gives the solution:
\bea r=2M\,\ex{\l\big(\w-{1/\w}\big)}\ ;& \quad&\quad
A=\l\w\ ;\qquad\nn\\ 8\pi\r={3\over\w^2(2M)^2}\qquad;& \quad&
B={\w^3\over\l(1+\w^2)^2} \,,\eea 
 holding as long as
\be \left|{r\over 2M}-1\right|\ll \l^{\ffract23}\,. \ee 
 
This overlaps with region~(5), where integrating the equations yields
\bea &&  {r\over 2M}= (1+5P)^{-\ffract15}\ \ ; \qquad\quad 
A={\l^2\over P} \left({2M\over r}\right)^6\,;\nn\\ &&B={\l^2\over
P^3}\left({2M\over r}\right)^{14}\ \ ;\qquad\quad
8\pi\r={3P^2\over\l^2(2M)^2}
\left({r\over2M}\right)^{12}\,.\label{er5}\eea 
 
Finally, in region~(6) we have ${r\over2M}\ll 1$, and our solution
simplifies into \be 
P={1\over5}\left({2M\over r}\right)^5\ ;\qquad A=5\l^2\,{2M\over r}\ ; 
\qquad B=125\l^2\,{r\over2M} \,. \ee  
 
The distance (in Planck units) between the point $r=2M$ (more or 
less in the middle of
region~(3)) and the origin at $r=0$ follows from Eqs.~(3.8):
\be  \int_0^{2M}\sqrt{B(r)}\dd r\ =\   \int_0^12M\sqrt\l{\dd\w\over\sqrt\w}\ =\ 
4M\sqrt\l\ =\  \sqrt{M\over 3}\left({N\over 5\pi}\right)^{\ffract14}\,. \ee    
 
In contrast, region~(5) is small (in spite of the fact that $r$ runs
from $0$ to nearly $2M$), its geodesic length being proportional to an
even smaller power of the black hole mass $M$.
With $y=(2M/r)^5=5P$, the metric here can be written as
 
\be \dd s^2\ =\ -{5\l^2 y^{\ffract65}\over y-1}\dd t^2\ +\ 5(2M\l)^2{y^{\ffract25}
\over(y-1)^3}\dd y^2\ +\ y^{-\ffract25}\dd\tl x^2\,. \ee  
Substituting \eqn{lambda}, and rescaling the time $t$, shows that this metric
will be universal for all black holes. It is the metric of a horizon,
as seen by a `Rindler observer'.

The gravitational potential $A(r)$ takes an absolute minimum inside
this region~(5):
\be \part Ar=0\quad\ra\quad  y=6\quad ,\quad {r\over2M}={1\over\root
{\scriptstyle 5}\of 6}\,,  \ee 
and here, the matter density takes the extreme value
\be \r^{\rm extr}={3(y-1)^2\over200\pi(2M\l)^2}y^{-\ffract{12}5}=
{5\cdot 6^{\fract35}\over N}\,. \ee 
Observe, that this is inversely proportional to the number of fields $N$
contributing to the Hawking radiation.
 
\secnum{The entropy}
According to local observers,
the entropy density $s$ of matter at any
$\k$ value, is 
\be s=\b(r)(1+\k)\r\,. \ee  Here, however, one has to substitute the
locally observed temperature, which, due to redshift, is given by
\be \b(r)=\sqrt{A(r)}\,\b\,, \ee  where $\b$ is the inverse
temperature as experienced by the distant observer.

The entropy for general $\k$ was also calculated by the author, but it 
also can be deduced from Ref. \cite{ZP}, who use the coefficient $\g=\k+1$, 
or $n=1+{1\over\k}$. The main contribution comes from region (5). As Eqs~\eqn{er5}
were computed for the case $\k=\fract13$, we first give the entropy for
that case:
\be {S\over\SS}={2\pi^2N\over45\b^3}\,{2M\over 5\l^2}\int_1^\infty
 {\dd y\over y^2}\,,\ee
where $\SS$ is the area of the horizon. With the value \eqn{lambda} for
$\l$, one obtains
\be S/\SS\ =\ \fract25\ ,\ee
instead of Hawking's value $\fract14$. Now, for the more general equation of
state, Eq.~\eqn{kappa}, this calculation can be repeated (see the next
section), with the result
\be {S\over\SS}={\k+1\over 7\k+1}\,,\label{kentropy} \ee 
and this equals the desired value $\ffract14$ if $\k=1$.
It is difficult to imagine ordinary matter with such a high $\k$ value. 
Free massless fields generate the entropy density
\be  s=C\,N\,T^3\ ,\label{freeentropy} \ee 
where $N$ is the number of non-interacting field species. $C$ is a universal
constant. It appears that, at very high temperature, this should be replaced by
\be  s=C\,T^{1/\k}\ .\label{kdensity} \ee 
A striking feature of the result of Eq.~\eqn{kentropy} is the independence on $N$. 
But, if $N$ is made
to depend strongly on temperature, this does affect the equation of
state. Now copare Eqs~\eqn{freeentropy} and \eqn{kdensity}. 
If a strong kind of unification takes place at Planckian
temperatures, such that, at those temperatures $N$ suddenly decreases
strongly, one could imagine an effective increase of $\k$ beyond the
canonical value $\ffract13$. It is more likely, however, that this argument
is still far too naive, and that our approach must merely be seen as
a rough approximation. An error of 60\% is perhaps not so bad. On the other hand,
it is tempting to speculate that the case $\k\ra1$ has physical significance. 
This is a highly peculiar case. The total entropy receives its main contribution
from a very tiny region in space where the matter density reaches values 
diverging exponentially with the small parameter $\l$. 
 
\secnum{Summary of the cases for $0<\k<1$, and the case $\k=1$} 
  
For general $\k$, Eqs.~\eqn{asympt} and \eqn{XY} are replaced by
\be 8\pi\r={\l^2\over \k(2M)^2}A\up{-{1+\k\over2\k}}\quad ,\qquad 
X=A\,r\up{-{4\k\over1+\k}}\,. \ee 
  It is convenient to define
\bea P={(2M)^2\over\l^2}\,{X\up{1+\k\over2\k}\over Y}\ ;\qquad
Q\,&=&\,1/Y\,.\eea
The field equations \eqn{PQ} become
\be {\dd P\over\dd L}={3\k+1\over2\k}\,{P\over Q}-{7\k+1\over2\k}P
-{1-\k\over2\k}\ , \qquad 
{\dd Q\over\dd L}=1-Q-{Q\over\k P}\, .\ee 

The solution of these equations can be obtained along the same lines as
above \cite{ZP}. The most important is the result in region~5:
\be \dd s^2=-a\,(y-1)\up{{-2\k\over 1-\k}}y\up{4\k(1+3\k)\over(1-\k)
(7\k+1)}\,\dd t^2 +b\,(y-1)\up{-2\over 1-\k}y\up{4\k(1+5\k)\over(1-\k)
(7\k+1)}\,\dd y^2 +y\up{-4\k\over 7\k+1}\,\dd\tl x^2\,\label{kdep}\ee where
$\dd\tl x^2$ stands for $(2M)^2\dd\W^2$, and
\be y=\left({2M\over r}\right)\up{7\k+1\over2\k}\ ;
\qquad a=\l\up{4\k\over1-\k}\left({1-\k\over7\k+1}\right)\up{-2\k\over1-\k}\ ;
\qquad b={4\k^2(7\k+1)\up{2\k\over1-\k}\over(1-\k)\up{2\over1-\k}}\,\l\up{4\k\over
1-\k}(2M)^2\ ,\label{kmetric}\ee 
and the entropy density turns out to be
\be s(r)={\l^2\b\over8\pi\k(2M)^2}(1+\k)\left(A(r)\right)\up{-{1\over2\k}}\,,\ee
where $A(r)$ stands for the term in front of $\dd t^2$ in Eq.~\eqn{kdep}.

The case $\k=1$ is particularly delicate \cite{GtH4}. Here we just quote the
result. We now find 8 regions, and of these, region~7 (which includes the
origin $r=0$) contains the most
interesting physics. The $\l$ dependence turns into an exponential one:
  
\be A=\l^\fract72\left({2M\over r}\right)\ex{{1\over\l^2}\left[\big({r\over2M}
\big)^4-1\right]}\ ,\qquad
B=\l^{-\half}\left({r\over2M}\right)\ex{{1\over\l^2}\left[
\big({r\over2M}\big)^4-1\right]}\,. \ee 
The entropy density is
\be s={\l^2\b\over8\pi\k(2M)^2}(1+\k)A\up{-{1\over2\k}}\,, \ee 
so that  
\be {S\over\SS}=\int_0^{2M}\left({r\over2M}\right)^3{\dd r\over2M}=\ffract14\,.
\ee

\end{document}